\documentclass{article}
\usepackage[english]{babel}

\usepackage[letterpaper,top=2cm,bottom=2cm,left=3cm,right=3cm,marginparwidth=1.75cm]{geometry}

\usepackage[colorlinks=true, allcolors=blue]{hyperref}

\usepackage{cite}
\usepackage{amsmath,amssymb,amsfonts}
\usepackage{algorithmic}
\usepackage{graphicx}
\usepackage{textcomp}

\title{Spatial Sampling of Hemispherical Arrays for Three-Dimensional Photoacoustic Computed Tomography}

\author{Wanqing Zhang$^1$, Hengyue Zhu$^1$, and Yide Zhang$^{1, 2, *}$}
\date{}
\begin{document}
\maketitle

\thanks{
$^1$ Department of Electrical, Computer and Energy Engineering, University of Colorado Boulder, Boulder, CO 80309 USA 

$^2$ Biomedical Engineering Program, University of Colorado Boulder, Boulder, CO 80309 USA 

$^*$ Corresponding author: Yide Zhang. E-mail: Yide.Zhang@colorado.edu}

\vspace{10pt}

\textbf{Abstract:} Three-dimensional (3D) photoacoustic computed tomography (PACT) is a powerful noninvasive biomedical imaging modality that provides volumetric data for structural and functional assessment \textit{in vivo}. To maximize angular coverage and mitigate limited-view artifacts, modern 3D PACT systems frequently employ hemispherical transducer arrays. While substantial effort has been devoted to improving image quality through post-processing algorithms, the intrinsic impact of the hardware array layout on the baseline image quality remains underexplored. In this study, we systematically investigate how the spatial sampling characteristics of hemispherical array distributions affect imaging performance under fixed hardware constraints. We propose a uniform-spacing sampling criterion to generate three representative array distributions and evaluate their performance using quantitative metrics across static, noise-perturbed, reduced-element, and rotational acquisition scenarios. Across these diverse testing regimes, the Fibonacci distribution consistently demonstrates superior structural robustness and more globally balanced reconstruction quality, a result we attribute to its highly isotropic sampling properties. These findings demonstrate that an optimal hardware-level sampling strategy is critical for maintaining global reconstruction stability. Ultimately, this framework establishes a rigorous quantitative methodology for benchmarking hemispherical array distributions and provides practical design guidance for the future development of 3D PACT systems.


\section{Introduction}
\label{sec:introduction}
Photoacoustic computed tomography (PACT) is an emerging imaging modality that uniquely combines the biochemical specificity of optical absorption with the deep-tissue penetration of ultrasound~\cite{wang_photoacoustic_2012}, which has advanced in both pre-clinical and clinical studies~\cite{li_single-impulse_2017,na_massively_2021,lin_single-breath-hold_2018}. By utilizing wide-field illumination and an ultrasonic transducer (UST) array to acquire signals over a large field of view (FOV), PACT avoids point-by-point scanning and enables fast monitoring of physiological activities~\cite{wang_practical_2016}. To accurately reconstruct the position and shape of optical absorbers in a three-dimensional (3D) space, the ideal detection strategy must capture the complete spherical photoacoustic (PA) wavefront~\cite{tian_spatial_2021}. Consequently, PACT systems are typically designed to approximate spherical detection geometries that satisfy reconstruction assumptions. Hardware implementations have progressed from linear~\cite{park_three-dimensional_2025}, half-ring~\cite{huang_high-speed_2025}, and full-ring arrays~\cite{li_single-impulse_2017}. Whereas these systems mainly support two-dimensional (2D) cross-sectional imaging, 3D imaging with these systems relies on synthetic-aperture acquisition~\cite{chen_photoacoustic_2024}, often resulting in non-isotropic resolution~\cite{miao_multiview_2025}. To overcome this limit, hemispherical arrays have been developed to provide wide angular coverage, enabling single-shot 3D imaging without the need for synthetic apertures~\cite{lin_high-speed_2021}.

Existing hemispherical arrays can be grouped into three categories: latitude-longitude distribution (LLD), in which USTs are arranged along the meridians and implemented either by direct fabrication or by rotating arc-shaped or half-ring arrays to synthesize a hemispherical surface~\cite{na_massively_2021, zhang_rotational_2026}; concentric-ring distribution (CRD), in which the hemisphere is partitioned into latitude-dependent rings with uniform circumferential spacing within each ring~\cite{gottschalk_rapid_2019, choi_deep_2023}; and Fibonacci distribution (FBD), in which USTs are uniformly distributed along the axial direction while azimuthal positions advance by a golden-angle increment~\cite{wang_cross-regional_2025, nagae_real-time_2019}. 

Reconstruction artifacts are still unavoidable in hemispherical array PACT because the detection geometry inherently covers less than the full 4$\pi$ solid angle and is further limited by a finite number of USTs with non-negligible physical size. To mitigate artifacts, prior studies have explored algorithmic solutions, including model-based reconstruction methods~\cite{dean-ben_accurate_2012}, deep-learning-based approaches~\cite{allman_photoacoustic_2018}, and anti-aliasing signal processing techniques~\cite{hu_location-dependent_2023}. From a hardware perspective, common interventions involve mechanically rotating the array to synthetically construct a virtual array with a higher effective sampling density~\cite{wang_cross-regional_2025}, or introducing acoustic reflectors to artificially increase angular coverage~\cite{sun_full-view_2024}.

However, a critical research gap remains: can specific UST array distributions intrinsically encode more information during data acquisition through a more effective sampling strategy, thereby improving baseline reconstruction quality? This gap arises from a common limitation in the existing literature, where most hemispherical array designs are described only qualitatively as being “evenly distributed”. The lack of a rigorous quantitative design methodology makes it difficult to isolate the effect of array geometry on imaging performance. Therefore, under identical constraints, such as a fixed number of USTs and a consistent hemispherical aperture, the specific impact of array distribution on spatial aliasing and reconstruction artifacts has not been systematically analyzed.

To address this gap, we propose a comprehensive array sampling strategy based on a uniform-spacing criterion to construct three representative hemispherical distributions. We quantitatively compare their intrinsic uniformity and periodicity and evaluate their baseline imaging performance through numerical simulations with a standardized reconstruction algorithm. Spatial resolution and usable FOV are measured with a point target, and the analysis is subsequently extended to vascular phantom imaging. To closely approximate realistic experimental conditions, we incorporate noise and the experimentally measured effective system impulse response (ESIR) of a physical UST into the simulated signals, enabling a systematic assessment of robustness against hardware imperfections and environmental noise. Finally, we investigate the performance discrepancies among these distributions under dynamic acquisition schemes. Together, this work transforms hemispherical array distribution design from an empirical practice toward a rigorous quantitative framework. By enabling predictive evaluation at the hardware design stage, we provide a robust pipeline for the future development of 3D PACT systems.

\section{Method}
\label{sec:method}
\subsection{Sampling strategy based on the uniform-spacing criterion}
\begin{figure}[!ht] 
\centerline{\includegraphics[width=0.7\columnwidth]{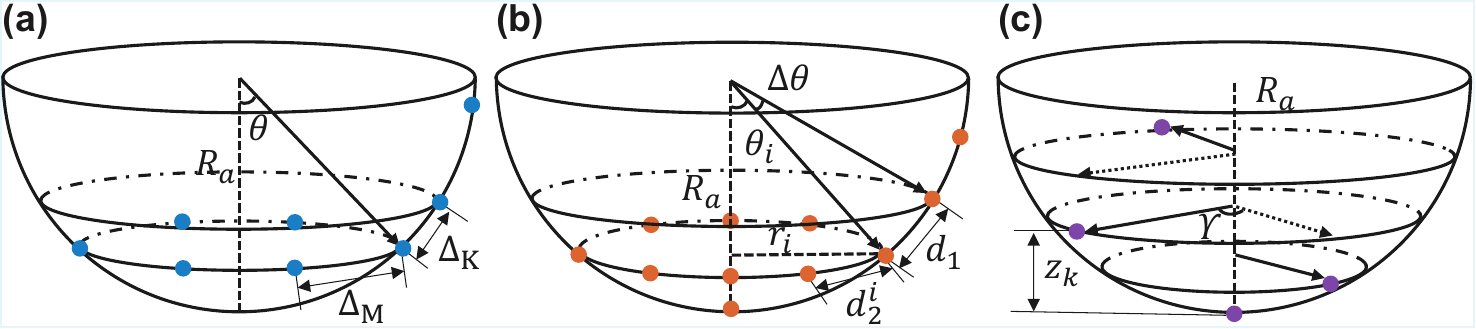}} 
\caption{Schematic diagram of the distribution. (a) LLD. (b) CRD. (c) FBD.} 
\label{distribution_demo} 
\end{figure}

To model a hemispherical array, we consider three key design parameters: the number of elements $N_e$, the hemispherical array radius $R_a$, and the element diameter $d_e$. In practice, $N_e$ is often restricted to $2^n$ due to the data acquisition channel count, while $R_a$ and $d_e$ are finite. Under these physical constraints, the sampling strategy can be formulated as seeking a distribution that achieves uniform spacing, in which each element corresponds to approximately equal surface area on the hemispherical surface. Following commonly adopted hemispherical array design paradigms in the literature, we construct three representative distributions using this uniform-spacing criterion and treat them as standardized candidates for quantitative comparison. The geometric construction is illustrated in Fig.~\ref{distribution_demo}, and the detailed construction procedure is described below.
\subsubsection{Latitude-longitude distribution}

The LLD can be parameterized by a factor pair $(K,M)$ of $N_e $: 
\begin{equation}
N_e = K M, 
\qquad K, M \in \mathbb{Z}^+.
\label{eq:Ne_relation}
\end{equation}
Here, $M$ denotes the number of meridians and $K$ denotes the number of points per meridian. Let $\theta$ denote the polar angle of the first parallel. With $K$ points uniformly distributed along each meridian from $\theta$ to the equator $(\pi/2)$, the longitudinal spacing is given by
\begin{equation}
\Delta_K = \frac{R_a \left( {\pi}/{2} - \theta \right)}{K - 1}.
\label{eq:longitudinal_spacing}
\end{equation}
Since $d_e \ll R_a$, the local distances on the sphere can be approximated by Euclidean distances. Therefore, the spacing constraint requires $\Delta_K \ge d_e$.
The most restrictive meridional constraint occurs on the first parallel, whose meridional spacing requires
\begin{equation}
\Delta_M = 2\pi R_a \sin\theta/M\ge d_e.
\label{eq:meridional_spacing}
\end{equation}
From Eq.~\eqref{eq:meridional_spacing}, the minimum polar angle is obtained as 
\begin{equation}
\theta_{\min}(K,M)
=
\arcsin\!\left(
{M d_e}/{(2\pi R_a)}
\right).
\end{equation}
The longitudinal constraint is then evaluated. If $\Delta_K(\theta_{\min},K) \ge d_e$, the pair $(K,M)$ is included in the feasible set $\mathcal{F}$. The optimal configuration is selected according to the uniform-spacing criterion: 
\begin{equation}
(K_{\mathrm{opt}}, M_{\mathrm{opt}})
=
\arg\min_{(K,M)\in\mathcal{F}} |K - M|,
\end{equation}
\begin{equation}
\theta_{\mathrm{opt}}
=
\theta_{\min}(K_{\mathrm{opt}},M_{\mathrm{opt}}).
\end{equation}
Balancing $K$ and $M$ in this manner reduces the aspect ratio of the local sampling cells, allowing the resulting grid elements to approach a square-like geometry. The corresponding $\theta_{\mathrm{opt}}$ maximizes the achievable polar coverage under the prescribed spacing constraints.

\subsubsection{Concentric ring distribution}
For CRD, we seek a distribution that satisfies the uniform-spacing criterion by balancing the inter-ring spacing $d_1$ and the circumferential spacing $d_2$ within a ring. To this end, we introduce a characteristic distance $d_c$ as the target spacing between neighboring elements, such that $d_1 \approx d_c$ and $d_2 \approx d_c$.

The polar angle of the $i$-th ring is
\begin{equation}
\theta_i = i \Delta\theta, 
\qquad i = 1,2,\dots,M,
\label{eq:theta_i}
\end{equation}
where $\Delta\theta = {d_c}/{R_a}$ and $M = \mathrm{floor}\left\lfloor {\pi}/{(2\Delta\theta)} \right\rfloor$. The corresponding inter-ring spacing on the spherical surface is $d_1 = R_a \Delta\theta = d_c$. The pole ($\theta = 0$) is treated as a single element.
At polar angle $\theta_i$, the radius of the ring is $r_i = R_a \sin\theta_i$, and the corresponding circumference is $C_i = 2\pi R_a \sin\theta_i$. 

The number of elements on the $i$-th ring is
\begin{equation}
n_i = \mathrm{round}\!\left( {C_i}/{d_c} \right),
\label{eq:ni}
\end{equation}
which leads to the actual circumferential spacing $d_2^i = C_i / n_i$. Minimum spacing constraints $d_1\ge d_e$ and $d_2^i \ge d_e$ are enforced whenever applicable.

The total number of elements implied by $d_c$ is
\begin{equation}
N_{\mathrm{total}}(d_c)
=
1 + \sum_{i=1}^{M} n_i.
\label{eq:Ncalc}
\end{equation}
A larger $d_c$ produces fewer rings and fewer elements per ring. Therefore, $N_{\mathrm{total}}(d_c)$ is approximately monotonically decreasing with respect to $d_c$, despite the discretization introduced by $\mathrm{round}(\cdot)$ and $\mathrm{floor} \lfloor \cdot \rfloor$.
We therefore apply a binary search to determine $d_c^\ast$ such that $N_{\mathrm{total}}(d_c^\ast)$ matches the desired channel count $N_e$ as closely as possible.
If the resulting distribution yields $N_{\mathrm{total}} \neq N_e$, we adjust the per-ring element counts while minimally perturbing spatial uniformity. Let $\Delta N = N_e - N_{\mathrm{total}}$. If $\Delta N > 0$, elements are added to the sparsest ring (largest $d_2^i$); if $\Delta N < 0$, elements are removed from the densest ring (smallest $d_2^i$), until $\Delta N = 0$.

\subsubsection{Fibonacci distribution}
The spherical Fibonacci lattice intrinsically provides an approximately uniform, equal-area 
distribution of sampling points on a sphere~\cite{swinbank_fibonacci_2006}. 
The construction combines an irrational golden-angle increment in azimuth 
to suppress periodicity and equal-area sampling in latitude via an even increment in the axial coordinate. Indexing $N_e$ transducers by $k = 0,1,\dots,N_e - 1$, the $k$-th element is uniquely given by
\begin{equation}
z_k = R_a({k + 0.5})/{N_e},
\label{eq:zk}
\end{equation}
\begin{equation}
\phi_k = (\gamma k) \bmod (2\pi),
\label{eq:phik}
\end{equation}
where $\gamma$ is the golden angle, defined as $\gamma = {2\pi}/{\varphi}$, with the golden ratio $\varphi = \frac{1 + \sqrt{5}}{2}$. Here, $z_k$ and $\phi_k$ denote the axial coordinate and azimuthal angle of the $k$-th element, respectively.

\subsubsection{Evaluation metrics for distribution}
To quantify the uniformity of the distributions, we compute the Voronoi cell areas associated with each element 
on the hemispherical surface~\cite{swinbank_fibonacci_2006}. Define the coefficient of variation (CV) as $\mathrm{CV}={\sigma_A}/{\mu_A}$, where ${\mu_A}$ is the mean Voronoi area, and ${\sigma_A}$ is the corresponding standard deviation (STD). For a uniformly distributed point set, the Voronoi areas are expected to be nearly identical, resulting in CV approaching zero.

To characterize spatial periodicity and anisotropy, we analyze the tangent-plane power spectral density (PSD) of the projected point configuration. The spatial distribution of array elements within a small spherical neighborhood is projected onto its tangent plane, where the planar approximation is valid due to the limited angular extent~\cite{calabretta_representations_2002}.
The projection is represented as a point-density field
\begin{equation}
\rho(\mathbf{r}) 
= 
\sum_{j=1}^{N} 
\delta\!\left( \mathbf{r} - \mathbf{r}_j \right),
\label{eq:rho_definition}
\end{equation}
where $\mathbf{r}$ is the 2D position vector on the tangent plane, $\mathbf{r}_j$ is the projected position of the $j$-th element, and $\delta(\cdot)$ denotes the Dirac delta function. The 2D Fourier transform of $\rho(\mathbf{r})$ yields
\begin{equation}
\hat{\rho}(\mathbf{k})
=
\sum_{j=1}^{N}
e^{-i \mathbf{k} \cdot \mathbf{r}_j},
\label{eq:rho_hat}
\end{equation}
where $\mathbf{k}$ is the 2D spatial frequency vector. The corresponding tangent-plane PSD is defined as
\begin{equation}
S(\mathbf{k})
=
\frac{1}{N}
\left|
\hat{\rho}(\mathbf{k})
\right|^2,
\label{eq:structure_factor}
\end{equation}
which is equivalent to the structure factor commonly used in statistical physics to characterize periodicity and anisotropy of point configurations~\cite{hansen_distribution_2013}. The resulting PSD enables intuitive visualization of directionally periodic structures, such as ring-like or stripe-like spectral patterns.

\subsection{Numerical simulation and image reconstruction}
All numerical experiments were conducted using the k-Wave MATLAB toolbox~\cite{treeby_k-wave_2010}. 
To control the computational cost of 3D simulations while preserving the key 
characteristics of a hemispherical PACT configuration, 
the hemisphere radius $R_a$ was set to 50 mm. Each transducer element had a center frequency of 1 MHz and a diameter $d_e = 2\,\mathrm{mm}$. The number of elements is fixed at $N_e = 1024$. 
To approximate the spatial averaging effect of finite-sized elements, each transducer was sub-sampled using an 8-point pattern. 
An 80\% Gaussian filter was applied to the recorded signals to emulate the finite bandwidth response of the transducers.
The acoustic medium was set with the speed of sound $c = 1500\,\mathrm{m/s}$ and the density 
$\rho = 1000\,\mathrm{kg/m^3}$. 
Acoustic attenuation was included using a power-law model~\cite{treeby_modeling_2010} 
with parameters $\alpha_{\mathrm{coeff}} = 0.75$ and $\alpha_{\mathrm{power}} = 1.5$. Image reconstruction was performed using the universal back-projection (UBP) algorithm~\cite{xu_universal_2005} 
as a common baseline for all experiments to isolate the effect of array element distribution and ensure fair comparison. No additional post-processing was applied.

\subsection{Evaluation metrics for imaging performance}

To systematically evaluate imaging performance across different distributions, we adopt a multi-level assessment strategy. Spatial resolution is characterized using point-source experiments and reported as the full width at half maximum (FWHM) of the reconstructed point spread function (PSF) after Hilbert transform processing. For quantitative evaluation of vascular reconstruction fidelity, four metrics are employed: peak signal-to-noise ratio (PSNR), structural similarity index measure (SSIM)~\cite{zhou_wang_image_2004}, contrast~\cite{kempski_application_2020}, and high-frequency error norm (HFEN)~\cite{ravishankar_mr_2011}.

Let $I$ denote the reference image and $K$ the reconstructed image, each containing $N$ voxels. The PSNR is defined as
\begin{equation}
\mathrm{PSNR} =
10 \log_{10}
\left(
\frac{\max(I_i)^2}
{\frac{1}{N}\sum_{i=1}^{N}(I_i-K_i)^2}
\right).
\label{eq:psnr}
\end{equation}
Here, $I_i$ and $K_i$ denote the voxel intensities of the reference and reconstructed images, respectively. PSNR quantifies global voxel-wise fidelity.

The SSIM is computed as
\begin{equation}
\mathrm{SSIM}(I,K) = 
\frac{(2\mu_I \mu_K + C_1)(2\sigma_{IK} + C_2)}
{(\mu_I^2 + \mu_K^2 + C_1)(\sigma_I^2 + \sigma_K^2 + C_2)},
\label{eq:ssim}
\end{equation}
where $\mu_I$ and $\mu_K$ denote the local mean intensities, $\sigma_I^2$ and $\sigma_K^2$ are the local variances, and $\sigma_{IK}$ is the local covariance between $I$ and $K$. $C_1$ and $C_2$ are small constants introduced for numerical stability. SSIM evaluates structural similarity.

Image contrast (CONT) is defined as
\begin{equation}
\mathrm{CONT} = 
({\mu_{\mathrm{sig}} - \mu_{\mathrm{bg}}})/{\mu_{\mathrm{bg}}},
\label{eq:contrast}
\end{equation}
where $\mu_{\mathrm{sig}}$ and $\mu_{\mathrm{bg}}$ denote the mean intensities of the signal and background regions of interest.

The HFEN is defined as
\begin{equation}
\mathrm{HFEN} =
\left\| \mathrm{LoG}(I) - \mathrm{LoG}(K) \right\|_2,
\label{eq:hfen}
\end{equation}
where $\mathrm{LoG}(\cdot)$ denotes convolution with a Laplacian-of-Gaussian (LoG) filter. The LoG kernel is defined as
\begin{equation}
h_{\mathrm{LoG}}(\mathbf{x}) =
\nabla^2
\left(
\frac{1}{2\pi\sigma^2}
\exp\!\left(-\frac{\|\mathbf{x}\|^2}{2\sigma^2}\right)
\right),
\end{equation}
with $\sigma$ = 1.5 pixels in our implementation. HFEN emphasizes high-frequency differences and is therefore sensitive to spatial resolution and edge preservation.

\subsection{Experimental setup and effective system impulse response modeling}
Fabricating and experimentally characterizing all candidate transducer arrays is impractical. Instead, we incorporate the experimentally measured system response into the simulated signals for realistic bandwidth, phase response, and ring-down characteristics~\cite{park_comparison_2025}. Let $p_{\mathrm{ideal}}(t)$ denote the simulated pressure signal and $h_{\mathrm{ESIR}}(t)$ denote the ESIR. The effective signal used for image reconstruction is modeled as
\begin{equation}
p_{\mathrm{eff}}(t)
=
p_{\mathrm{ideal}}(t)
*
h_{\mathrm{ESIR}}(t),
\label{eq:peff}
\end{equation}
where
\begin{equation}
h_{\mathrm{ESIR}}(t)
=
h_{\mathrm{EIR}}(t)
*
h_{\mathrm{SIR}}(t)
*
h_{\mathrm{electronic}}(t).
\label{eq:heff}
\end{equation}
Here, $h_{\mathrm{EIR}}(t)$ is the electrical impulse response (EIR), $h_{\mathrm{SIR}}(t)$ is the spatial impulse response (SIR) and $h_{\mathrm{electronic}}(t)$ is the impulse response of additional system components such as amplification and electronic filtering devices. 

\begin{figure}[!h] 
\centerline{\includegraphics[width=0.7\columnwidth]{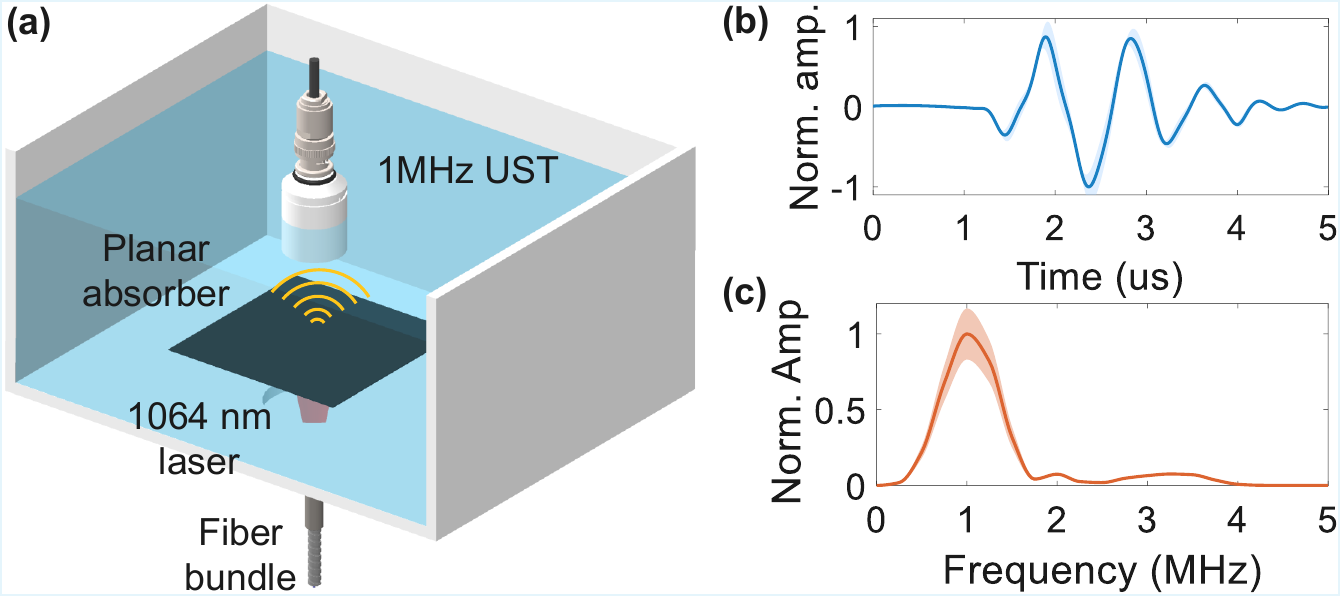}} 
\caption{Experimentally measured system response. (a) Experimental setup. (b) Normalized time-domain signal. (c) Corresponding frequency spectrum. The solid curves and shaded areas denote the mean and STD of the measured ESIR waveforms, respectively. } 
\label{EIRsetup} 
\end{figure}

The $h_{\mathrm{ESIR}}(t)$ is experimentally measured using a single-element UST and a planar absorber, capturing the overall acoustic-to-electrical response of the receive chain\cite{rosenthal_optoacoustic_2011,chowdhury_synthetic_2020}, as illustrated in Fig.~\ref{EIRsetup}(a). A 1 MHz unfocused UST (Olympus A303S) was immersed in water above the absorber, while a 1064 nm laser pulse (OPOTEK Opolucis C3020) was delivered through an optical fiber bundle and generated broadband PA signals. 
The measured $h_{\mathrm{ESIR}}(t)$ and its corresponding frequency spectrum are shown in Figs.~\ref{EIRsetup}(b) and (c), respectively.

\section{Results}
\subsection{Intrinsic characteristics and resolution analysis}
\begin{figure}[!ht] 
\centerline{\includegraphics[width=0.7\columnwidth]{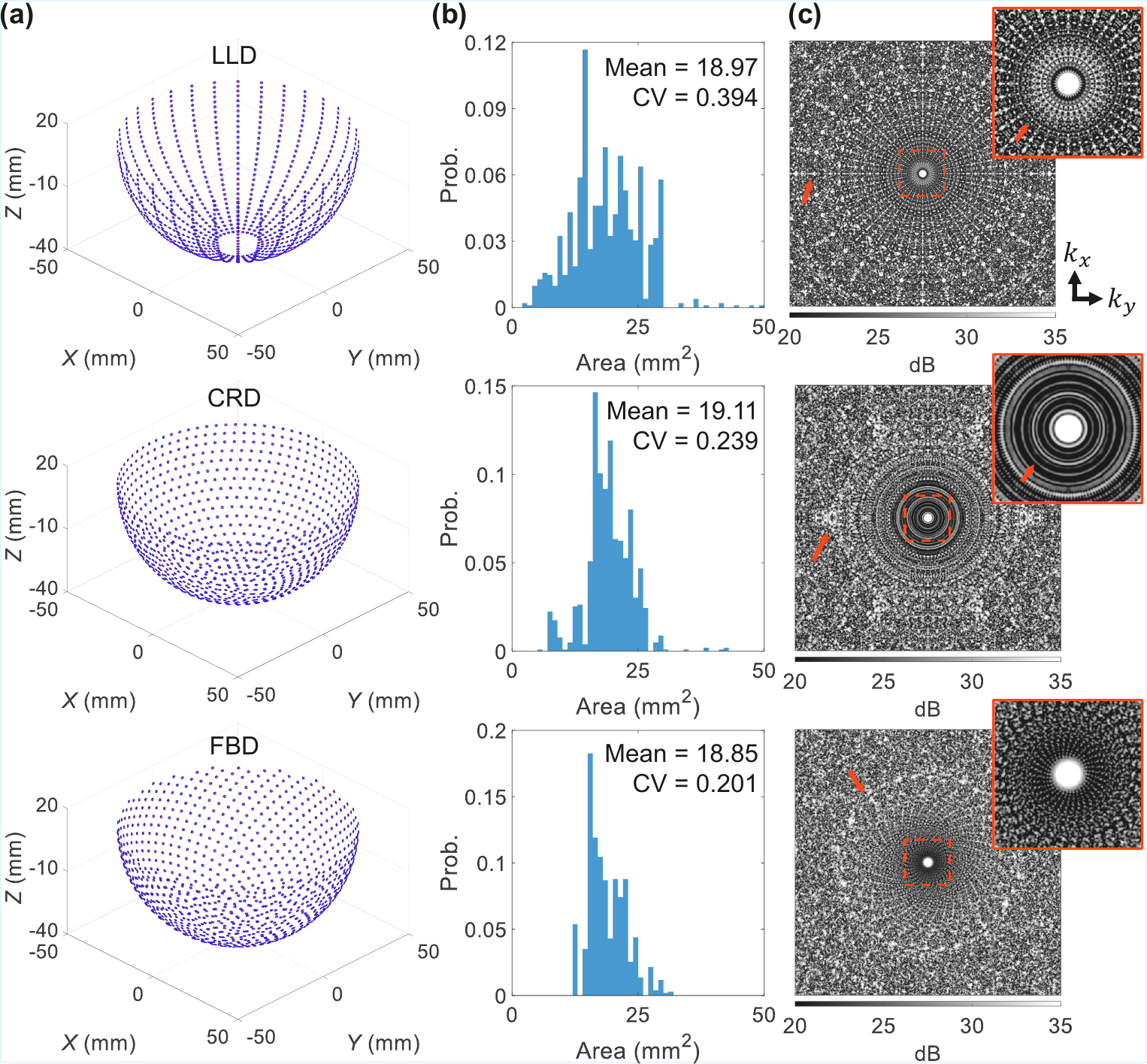}} 
\caption{Comparison of hemispherical distributions and their intrinsic spatial uniformity and periodicity. (a) 3D views of the three distributions. (b) Corresponding Voronoi cell area statistics. (c) Tangent-plane PSD.} 
\label{distribution_property} 
\end{figure}

Fig.~\ref{distribution_property}(a) illustrates the three hemispherical arrays generated using the proposed sampling strategy with the parameter settings described in the Methods section. 
For the LLD, the sampling strategy resulted in $(K,M)=(32,32)$ and $\theta_{\min} = 11.75^\circ$. For the CRD, the computed $d_1$ was 3.9~mm and the $d_2$ was 4.1 mm.

Among them, the FBD exhibits the highest spatial uniformity (the lowest CV of 0.201, Fig.~\ref{distribution_property}(b)). In contrast, the LLD exhibits the highest CV of 0.394, reflecting unequal surface coverage across elements. This non-uniformity arises from the excessive concentration of elements near the pole and sparse sampling around the equatorial region.

Each distribution exhibits a distinct and structured pattern in the tangent-plane PSD, as shown in Fig.~\ref{distribution_property}(c). As indicated by the red arrows, the LLD produces pronounced radial streaks, whereas the CRD generates hexagonal spectral patterns, indicating the periodicity and directional anisotropy. The FBD exhibits ring-like spectral features concentrated at specific spatial frequencies. In the low-frequency region, both the LLD and CRD display ripple-like circular patterns (insets of Fig.~\ref{distribution_property}(c)), indicating residual directional ordering.

\begin{figure}[!ht] 
\centerline{\includegraphics[width=0.7\columnwidth]{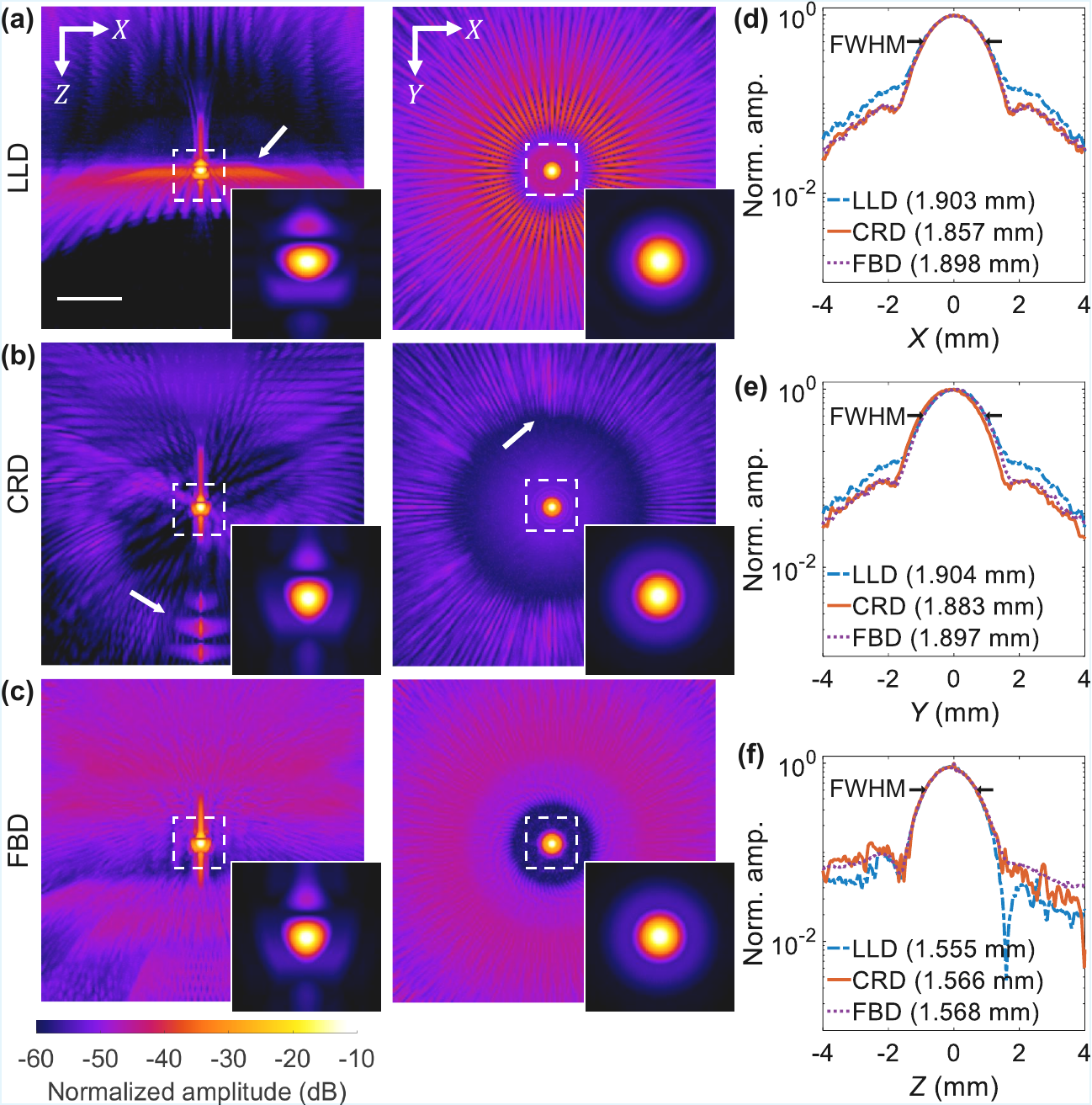}} 
\caption{Reconstruction of a central point target using different hemispherical distributions. (a) LLD. (b) CRD. (c) FBD. (d)–(f) PSF profile along \textit{X}, \textit{Y}, and \textit{Z} directions, respectively. The FWHMs are shown in the legends. The scale bar is 1 cm.} 
\label{PSF} 
\end{figure}

To evaluate the spatial resolution, a point target was placed at the geometric center of each hemisphere. Image reconstruction was performed using UBP over a 50 mm$^3$ volume (Figs.~\ref{PSF}(a)--(c)) to visualize the energy distribution in a large space, and over 8 mm$^3$ to visualize the detailed PSF profile (Figs.~\ref{PSF}(d)--(f)).

In the \textit{XY} plane, the LLD exhibits pronounced radial patterns that resemble the directional structures seen in the PSD analysis (Fig.~\ref{distribution_property}(c)). The CRD shows relatively clean reconstruction near the center, but structured intensity modulations appear at larger radial distances (the white arrows in Fig.~\ref{PSF}(b)). Along the \textit{Z} direction, the LLD shows severe artifact propagation (Fig.~\ref{PSF}(a)). The CRD generates localized bright spots near the image boundary at 26.5 mm, 33.5 mm, and 40.5 mm away from the hemisphere center. In comparison, the FBD exhibits comparatively uniform energy dispersion across the whole 3D space (Fig.~\ref{PSF}(c)). This isotropic behavior reduces the risk of direction-specific artifacts being misinterpreted as genuine anatomical structures, which is essential for \textit{in vivo} applications; without prior knowledge of the true anatomy, any structured artifact could easily be confused with actual biological features.

The CRD achieved a resolution of 1.857 mm in the \textit{X} and 1.883 mm in the \textit{Y}, indicating that the point source is distorted (Figs.~\ref{PSF}(d) and (e)). The LLD yielded the worst resolutions of 1.903 mm. All three distributions exhibit axial (\textit{Z}) FWHM values within approximately $\sim$1.6 mm (Fig.~\ref{PSF}(f)). However, this does not indicate an improvement in axial resolution. Instead, because of the incomplete hemispherical aperture along the \textit{Z} direction, the PSF is modified by the negative artifacts and spatial compression and exhibits distortion (see the first column of Fig.~\ref{PSF}). After applying the Hilbert transform, this distortion leads to artificially reduced FWHM values.

The usable FOV was evaluated by examining how the spatial resolution varies with source location. Point sources were placed along the axial direction with \textit{X}, \textit{Y} = 0 and \textit{Z} $\in$ [-37.5,\,27] mm, and along the radial direction with \textit{Y}, \textit{Z} = 0 and \textit{X} $\in$ [0,\,37.5] mm. 

\begin{figure}[!t] 
\centerline{\includegraphics[width=0.7\columnwidth]{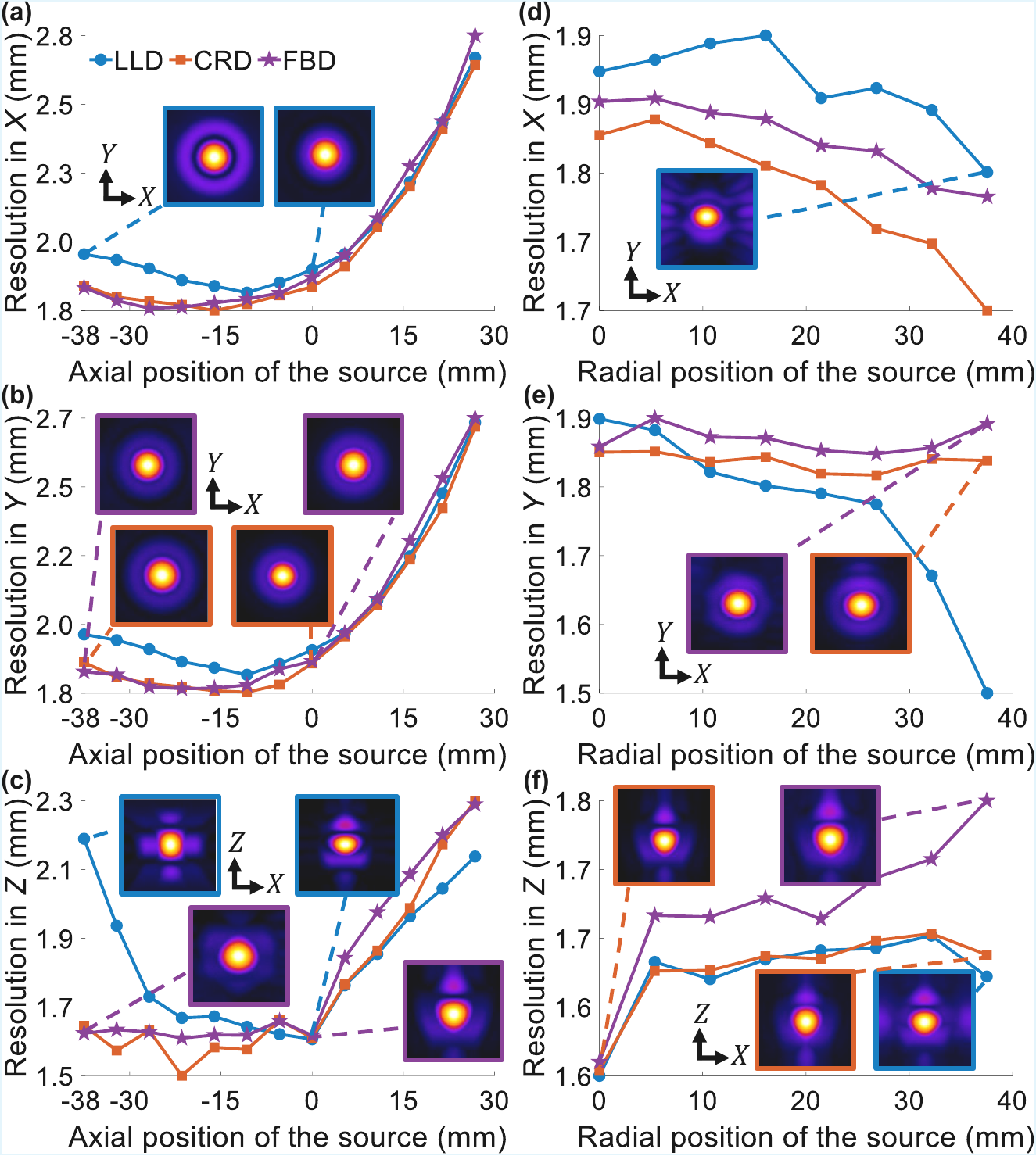}} 
\caption{Resolution variation with point source position for different hemispherical distributions. (a)–(c) Resolution measured along the axial scan. (d)–(f) Resolution measured along the radial scan. The insets are the reconstruction results of three distributions.} 
\label{FOV} 
\end{figure}

Figs.~\ref{FOV}(a)--(c) show the resolution variation along the axial direction. The LLD exhibits slight degradation in \textit{X} and \textit{Y} resolution at negative axial positions (from 1.90 mm to 1.96 mm), whereas the other two distributions maintain a relatively stable level of $\sim$1.85 mm. When the source moves outside the hemispherical aperture toward positive axial positions, the resolution of all distributions rapidly deteriorates to $\sim$2.70 mm. For \textit{Z} resolution, the LLD shows a sharp deterioration (as shown in the insets) at positions below -20 mm (up to 2.15 mm). In contrast, the FBD maintains a nearly constant \textit{Z} resolution of $\sim$1.6 mm. The CRD shows oscillatory variations, reflecting non-uniform spatial sampling.

Along the radial scan, the \textit{X} resolution decreases with radial distance for all distributions (Fig.~\ref{FOV}(d)). The \textit{Y} resolution remains relatively stable for the CRD and FBD, whereas the LLD shows noticeable degradation due to distortion of the reconstructed PSF at off-center locations, consistent with the previous analysis. At large radial distances (e.g., 37.5 mm), the LLD reconstruction becomes visibly distorted (insets of Fig.~\ref{FOV}(d)), while the FBD and CRD maintain more consistent shapes (insets of Fig.~\ref{FOV}(e)). For the \textit{Z} resolution, the LLD and CRD remain relatively stable at $\sim$1.65 mm, whereas the FBD shows a gradual increase with radial distance. However, the insets in Fig.~\ref{FOV}(f) show that the FBD preserves a nearly unchanged PSF shape with only slight elongation along \textit{Z}, whereas the CRD and the LLD reconstruction exhibits alterations in the artifact pattern. Overall, although the measured FWHM varies across positions, the FBD maintains a PSF profile comparable to that at the spherical center over a larger FOV. This suggests an expanded usable imaging region.

\subsection{Static imaging performance and noise robustness}

\begin{figure}[!h] 
\centerline{\includegraphics[width=0.5\columnwidth]{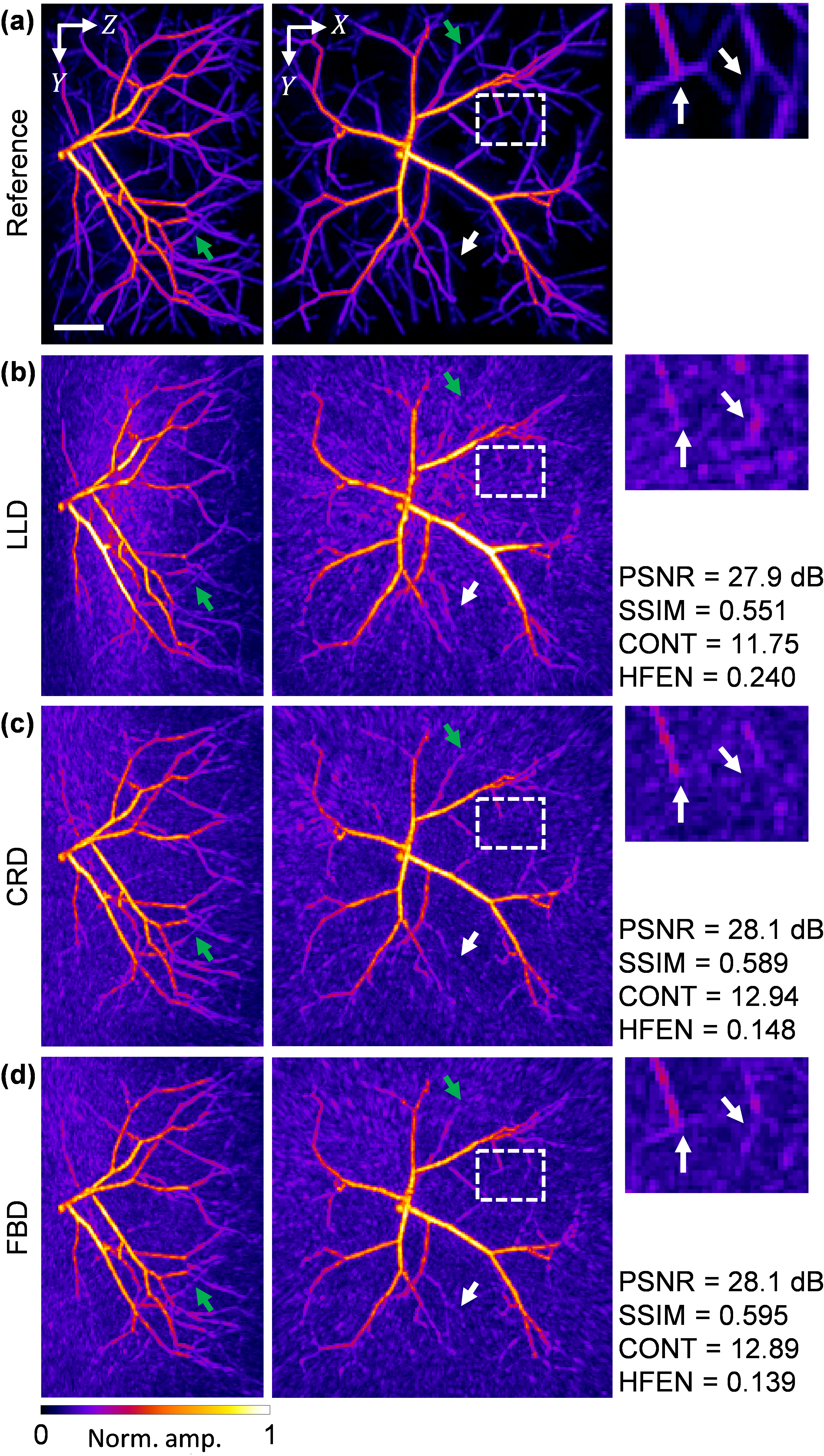}} 
\caption{Vascular phantom reconstruction. (a) Reference image. (b)--(d) Reconstruction images obtained using the LLD, CRD, and FBD, respectively. The scale bar is 1 cm.} 
\label{vessel} 
\end{figure}

The macroscopic imaging performance was evaluated using a vascular phantom~\cite{hamarneh_vascusynth_2010}, as shown in Fig.~\ref{vessel}. We used the same 1024-element array distributions and the parameter settings as in Fig.~\ref{distribution_property}. To establish a ground truth for benchmarking the 1024-element arrays, the reference was imaged with a dense CRD containing 20480 elements, which follows the proposed sampling strategy. Considering this dense array introduces element overlap, it establishes an upper bound of achievable image quality under hemispherical coverage.

The FBD achieves the generally best imaging quality. Quantitative evaluation indicates that the FBD yields the highest SSIM (the higher the better) of 0.595 and the lowest HFEN (the lower the better) of 0.139 values among three distributions, although with the same PSNR (the higher the better) and slightly lower CONT (the higher the better) values compared with the CRD. The LLD distribution performs substantially worse than the other distributions across all metrics. Visual inspection reveals that the CRD and FBD produce comparable reconstruction quality. However, the FBD preserves finer vascular structures more effectively (the dashed boxes in Fig.~\ref{vessel}). Direction-dependent behavior is observed for the CRD (the green arrows), where fine structures are preserved along certain orientations but degraded along others (the white arrows). This observation can be attributed to the anisotropic structured patterns of the CRD, which are identified in the tangent-plane PSD analysis (Fig.~\ref{distribution_property}(c)).

\begin{figure}[!t] 
\centerline{\includegraphics[width=0.5\columnwidth]{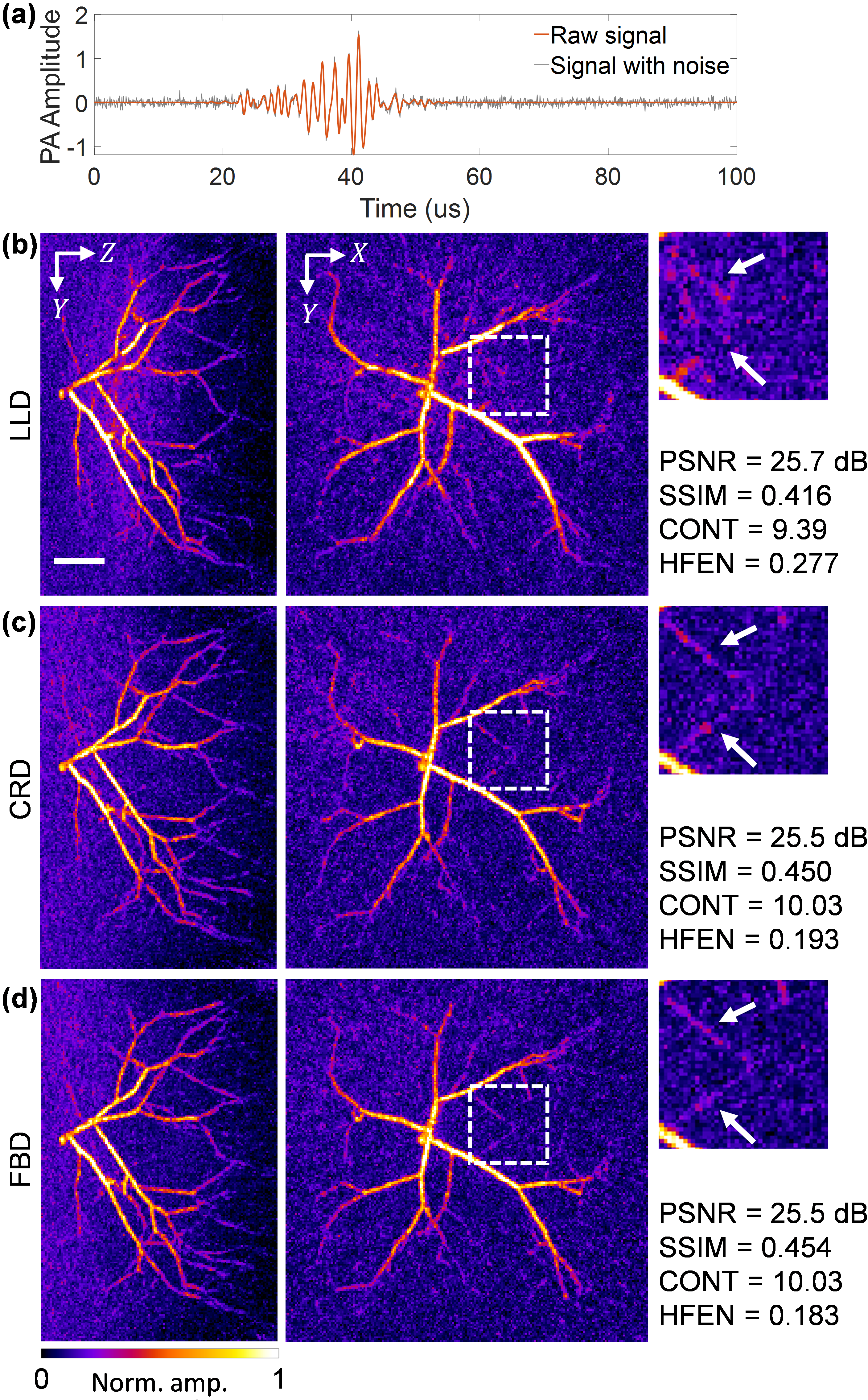}} 
\caption{Vascular phantom reconstruction with adding Poisson noise (peak-count=1000) and Gaussian noise (NSR=-15 dB). (a) Representative channel signals. (b)–(d) Reconstruction images obtained using the LLD, CRD, and FBD, respectively. The scale bar is 1 cm.} 
\label{noise1} 
\end{figure}

\begin{figure}[!t] 
\centerline{\includegraphics[width=0.5\columnwidth]{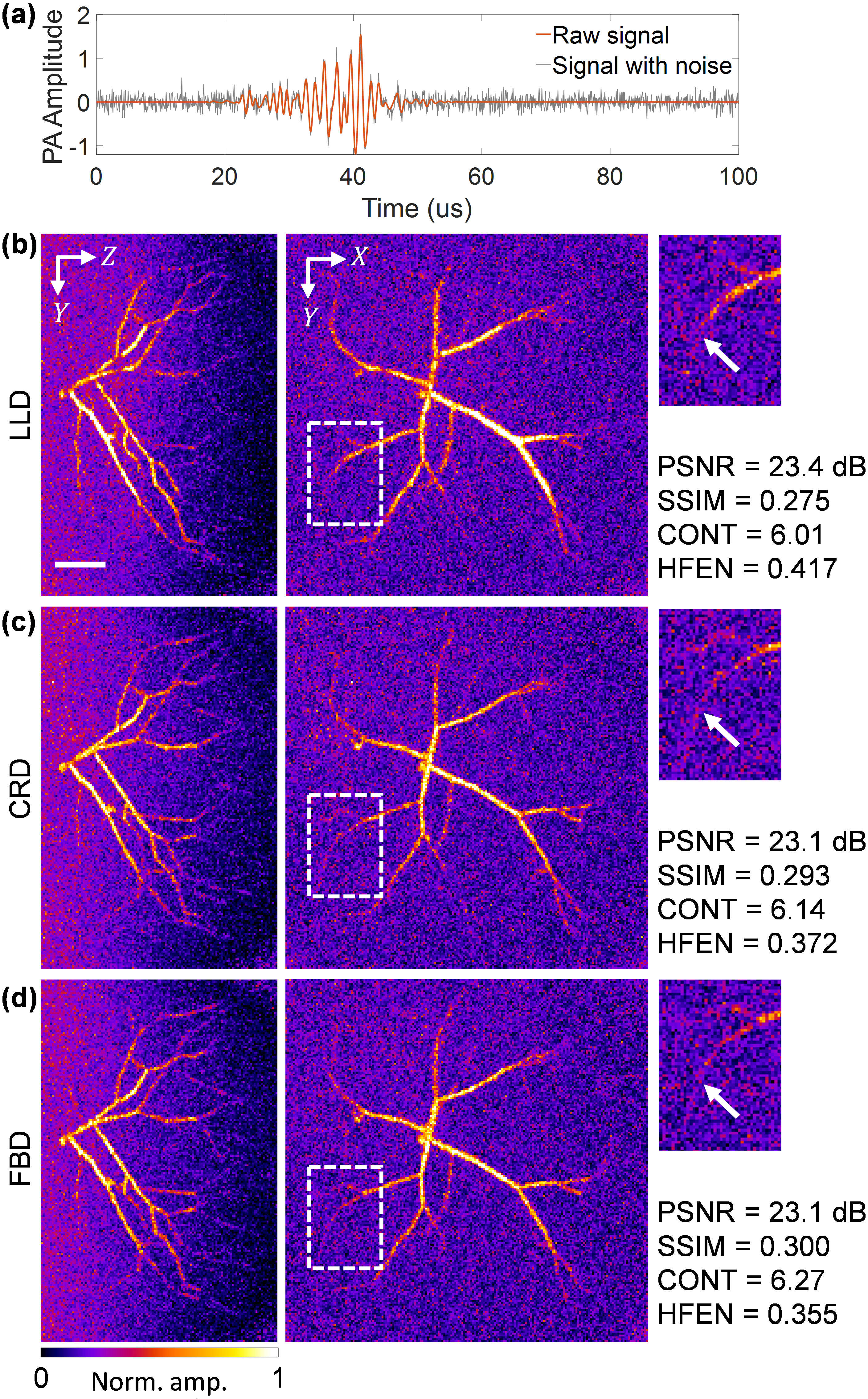}} 
\caption{Vascular phantom reconstruction with adding Poisson noise (peak-count=200) and Gaussian noise (NSR=-15 dB). (a) Representative channel signals. (b)–(d) Reconstruction images obtained using the LLD, CRD, and FBD, respectively. The scale bar is 1 cm.} 
\label{noise2} 
\end{figure}

We further evaluated the robustness of the three distributions under noisy conditions. To emulate experimental noise, for each channel, Poisson noise was first introduced with a peak photon count of 1000, followed by additive white Gaussian noise. The Gaussian noise level was controlled by specifying the noise-to-signal ratio (NSR), defined in the amplitude domain as $\mathrm{NSR} = {\sigma_g}/{\mathrm{RMS}(s)}$, where ${\mathrm{RMS}(s)}$ denotes the root mean square of the Poisson-corrupted signal and $\sigma_g$ is the STD of the Gaussian noise.
The NSR was set to -15 dB in Fig.~\ref{noise1}. Fig.~\ref{noise2} shows a more severe noise scenario: the peak photon count was reduced to 200 while keeping the NSR unchanged. The identical noise realization was applied to all three array distributions to ensure a fair and controlled comparison.


Even in the presence of noise, the FBD consistently delivers the best overall imaging quality, although its PSNR is slightly lower than that of the LLD. In Fig.~\ref{noise1}, the FBD achieves the highest SSIM value (0.454), the highest contrast (10.03), and the lowest HFEN value (0.183), indicating improved structural preservation and reduced high-frequency errors. Under the more severe Poisson noise condition (Fig.~\ref{noise2}), this observation remains consistent. In particular, at larger radial distances, it preserves more continuous and well-defined vascular structures (as indicated by the dashed boxes and white arrows). In contrast, although the LLD attains the highest PSNR (25.7 dB and 23.4 dB), visual inspection reveals noticeable structural discontinuities and locally missing branches, especially in the peripheral regions of the FOV. As a global intensity-based metric, PSNR is less sensitive to localized structural degradation and may therefore overestimate perceptual image quality in this context\cite{zhou_wang_mean_2009}.

\subsection{Imaging performance with experimentally measured system response}

\begin{figure} [!h] 
\centerline{\includegraphics[width=0.5\columnwidth]{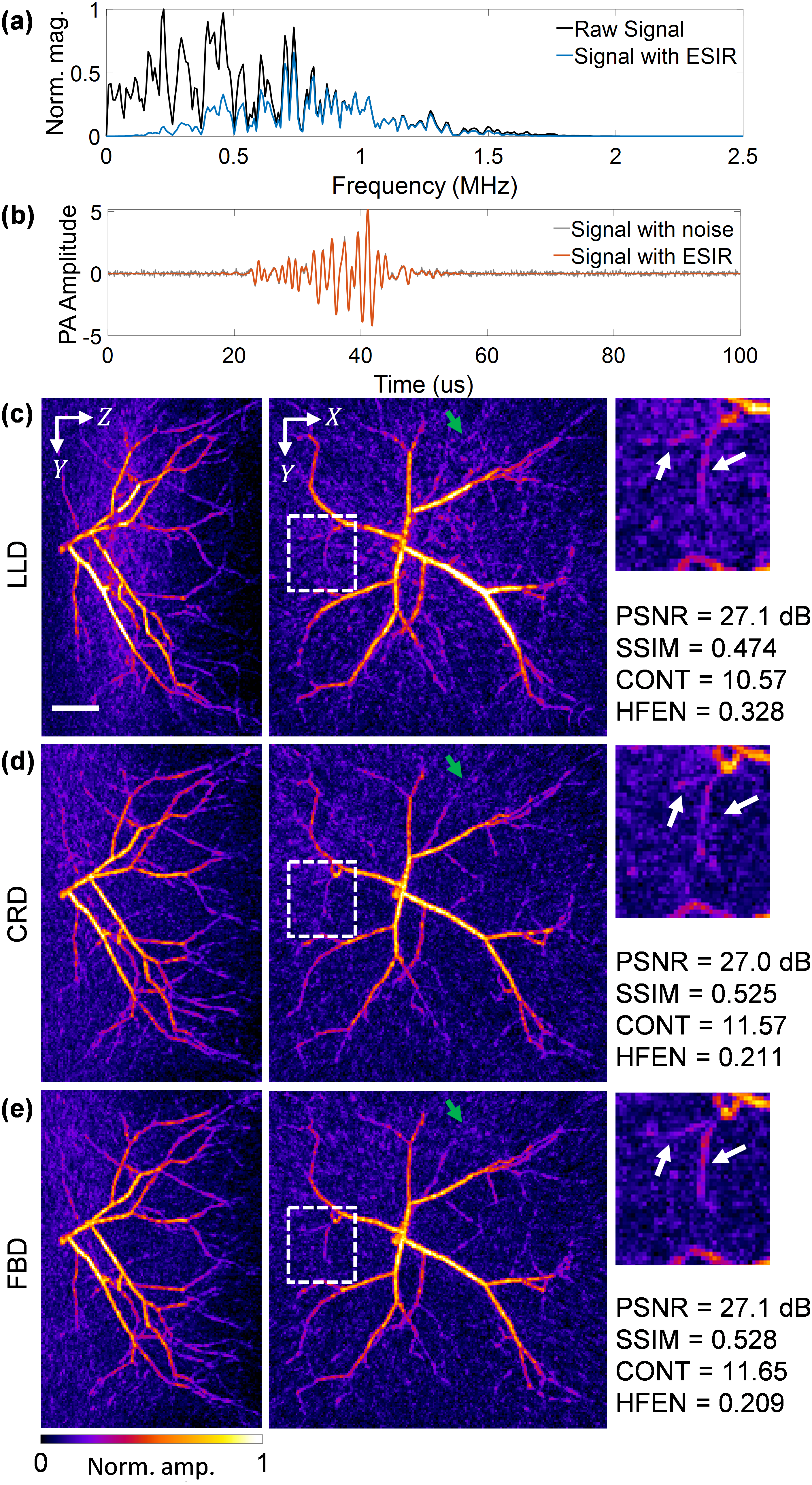}} 
\caption{Vascular phantom reconstruction with ESIR and Poisson and Gaussian noise. (a) Representative channel spectra before and after convolution with the ESIR. (b) A representative channel signal with the ESIR and noise. (c)–(e) Reconstruction images obtained using the LLD, CRD, and FBD, respectively. The scale bar is 1 cm.} 
\label{vesselESIR} 
\end{figure}

The impact of the experimentally measured ESIR on imaging performance was investigated. The average ESIR (Fig.~\ref{EIRsetup}(b)) was convolved with the simulated pressure signal of each sensor channel before reconstruction, resulting in Fig.~\ref{vesselESIR}(a). The averaged response suppresses a portion of the stochastic noise present in individual channels. Moreover, directly convolving the simulated signals with a single ESIR would implicitly assume identical responses for all array elements, whereas in practice each channel exhibits slight variations. To better approximate realistic measurement conditions and restore channel variability, we therefore introduced additional Poisson noise (peak count = 3000) and Gaussian noise (-30 dB NSR) after convolution (Fig.~\ref{vesselESIR}(b)). Because the ESIR inherently includes the SIR and EIR, each array element was modeled as a single point detector, and the 80\% bandwidth Gaussian filter was not applied. The reconstructed results are shown in Figs.~\ref{vesselESIR}(c)–(e). 
\begin{figure}[!h] 
\centerline{\includegraphics[width=0.5\columnwidth]{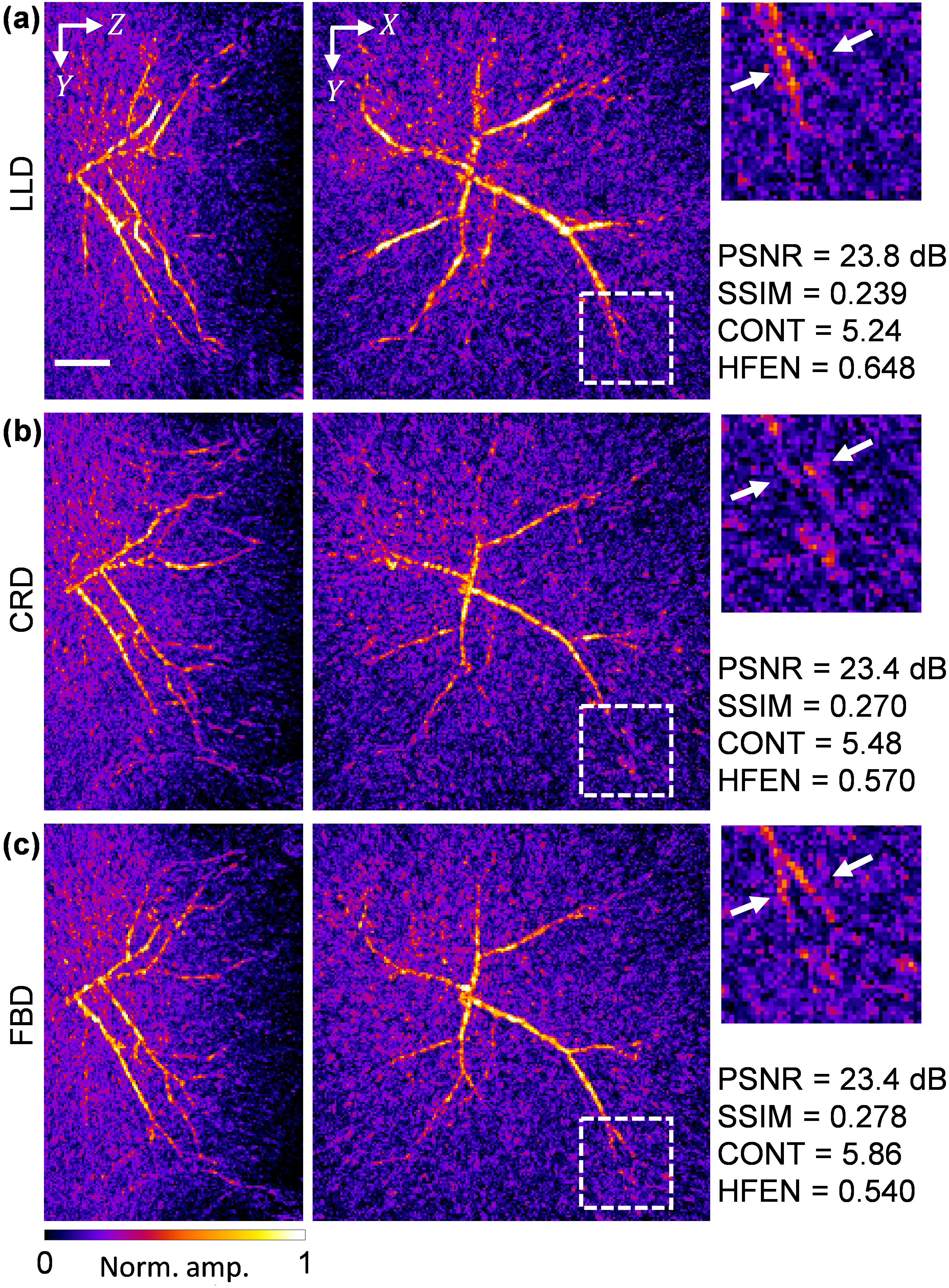}} 
\caption{Vascular phantom reconstruction with the ESIR and Poisson and Gaussian noise under 256-element distributions. (a)–(c) Reconstruction images obtained using the LLD, CRD, and FBD distributions, respectively. The scale bar is 1 cm.} 
\label{vessel256} 
\end{figure}
\begin{figure}[!ht] 
\centerline{\includegraphics[width=0.5\columnwidth]{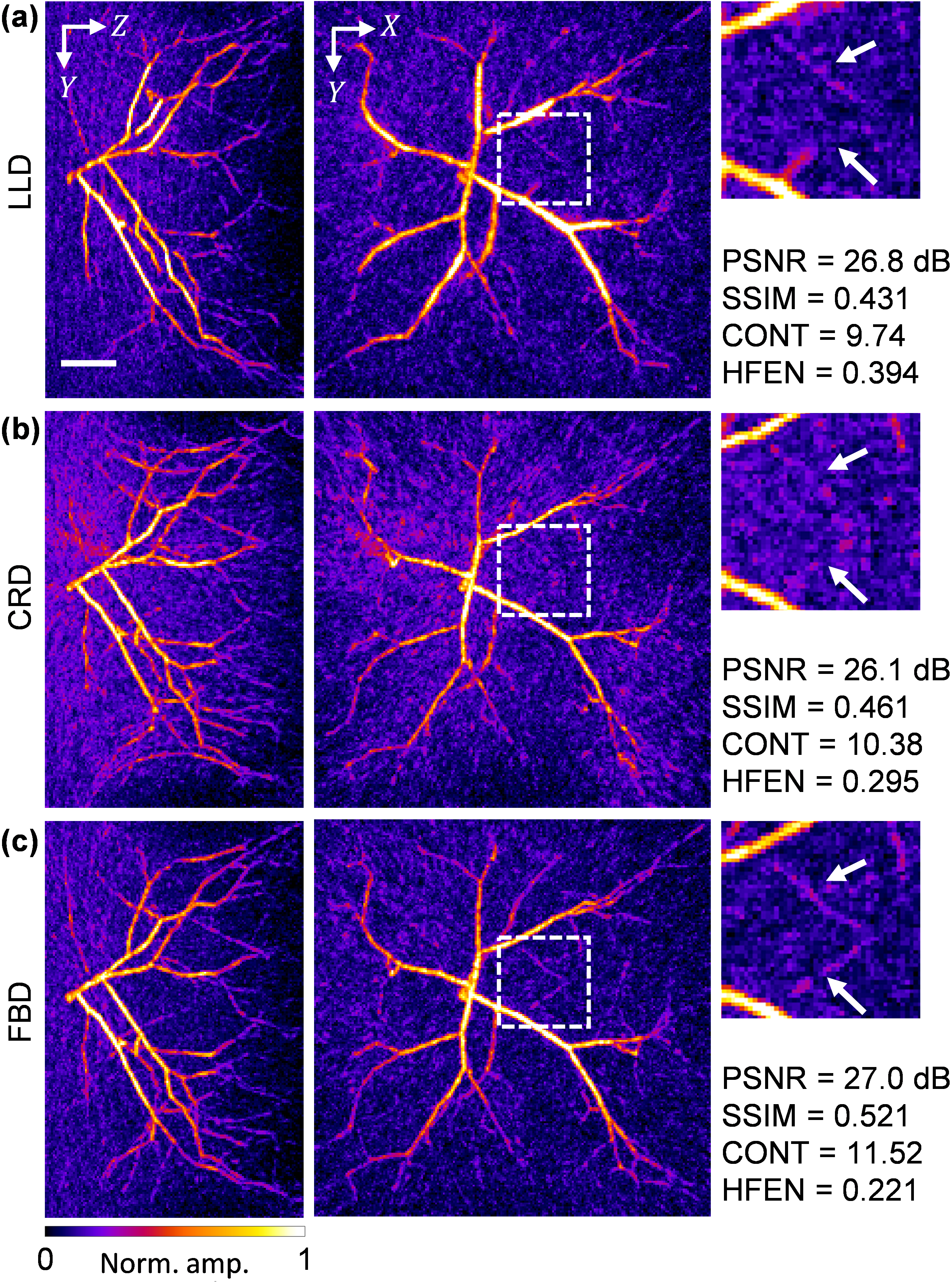}} 
\caption{Vascular phantom reconstruction with the ESIR and Poisson and Gaussian noise under virtual 1024-element distributions constructed by rotating the 256-element distributions. (a)–(c) Reconstruction images obtained using the LLD, CRD, and FBD, respectively. The scale bar is 1 cm.} 
\label{virtual1024} 
\end{figure}

Under the joint influence of realistic ESIR and noise contamination, the relative performance differences among distributions become more nuanced. While the CRD and FBD exhibit broadly comparable quantitative metrics, qualitative inspection reveals distinct structural behaviors. In the highlighted regions, the FBD maintains more continuous and coherent vascular branches (the dashed boxes and the white arrows in Fig.~\ref{vesselESIR}), whereas the CRD tends to produce locally enhanced structures (the green arrows), consistent with the results in Fig.~\ref{vessel}. Notably, although the LLD achieves competitive PSNR values, structural discontinuities and severe background artifacts remain apparent. This observation highlights that, under realistic bandwidth and noise constraints, appropriate sampling strategies become particularly important, as preserving global structural continuity under degradation is essential for biomedical imaging applications. 

To better reveal the robustness among distributions, the number of transducer elements was reduced from 1024 to 256. This configuration represents a sparser sampling on the hemispherical surface. As in the previous experiments, the simulated signals were convolved with the ESIR and the same level of noise. Under this sparse sampling condition, the FBD distribution still achieves the best imaging performance, as shown in Fig.~\ref{vessel256}. The quantitative evaluation is similar to the 1024 case: the CRD and FBD exhibit a similar level, whereas the FBD achieves a much finer and more continuous structure, as indicated by the dashed boxes and white arrows. Notably, unlike the 1024-element configuration, where performance differences were moderate, the gap among the FBD, the LLD, and the CRD becomes more pronounced when only 256 elements are used. We attribute the improved structural preservation of the FBD to its more balanced angular coverage under sparse sampling conditions. 

\subsection{Dynamic imaging performance}
The analysis is extended from static imaging to dynamic acquisition. Fig.~\ref{virtual1024} illustrates the reconstructed results on the same 256-element distributions (as in Fig.~\ref{vessel256}) rotated four times to form a virtual 1024-element configuration. This experiment aims to evaluate the best achievable performance for each distribution under rotational acquisition. For the LLD and the CRD, the rotation angles were determined based on the angular spacing between adjacent elements on the same ring. The angular interval between neighboring elements was divided into four equal segments, resulting in rotation steps of 5.625° for the LLD and 18° for the CRD. For the FBD, because each element occupies a distinct \textit{Z} position, a different strategy was applied. The hemisphere was partitioned into axial bands with a width equal to the physical element diameter $d_e$. Within each band, the smallest circumferential spacing between neighboring elements was identified and divided into four equal parts. This procedure resulted in a rotation step of 5.02°.

Under rotational acquisition, the FBD distribution achieves the best performance across all evaluation metrics, with PSNR = 27.0 dB, SSIM = 0.521, CONT = 11.52, and HFEN = 0.221. Visual inspection indicates reduced background artifacts and improved continuity of fine vascular structures compared with the other two distributions, as indicated by the dashed boxes and white arrows in Fig.~\ref{virtual1024}. Compared with the original static 1024-element configuration shown in Fig.~\ref{vesselESIR}, the rotated FBD distribution shows only minor differences, with PSNR decreasing by 0.1 dB, SSIM and CONT decreasing by 1.33\% and 1.12\%, respectively, and HFEN increasing by 5.74\%. These results indicate that rotational acquisition enables the FBD to achieve reconstruction quality nearly equivalent to that of the static full-array configuration. In contrast, rotational synthetic aperture provides limited improvement for the other two distributions. Relative to the original 1024-element configuration, the virtual 1024-element LLD exhibits noticeable degradation: PSNR decreases by 0.3 dB, SSIM and CONT drop by 9.1\% and 7.9\%, respectively, and HFEN increases by approximately 20.1\%. The virtual 1024-element CRD shows even stronger degradation, with PSNR decreasing by 0.9 dB, SSIM and CONT dropping by 12.2\% and 10.3\%, respectively, and HFEN increasing by nearly 40\%, indicating substantially elevated high-frequency reconstruction errors.

\section{Discussion and Conclusion}

This study demonstrates that array distribution design plays a critical role in determining imaging quality in hemispherical 3D PACT systems. Rather than relying solely on post-processing methods to improve image quality, a more effective approach is to optimize the sampling configuration in the pre-design stage, thereby maximizing the utilization of limited hardware resources during data acquisition. However, systematic investigations into what constitutes a “good” sampling design remain lacking. To our knowledge, this is the first study to benchmark LLD, CRD, and FBD distributions under identical conditions and to establish a quantitative link between geometric sampling characteristics and imaging performance. More importantly, we provide performance baselines for these three classes of distributions, offering practical design references for new hemispherical arrays. Given a specific sampling strategy and geometric constraints, our framework enables the generation of a fundamentally uniform hemispherical distribution and provides guidance on maximizing effective information capacity under a fixed hardware budget.

Under static imaging conditions, the FBD achieves the most balanced overall performance, providing spatially consistent reconstruction quality and enhanced robustness to noise (Figs.~\ref{noise1} and \ref{noise2}), particularly under sparse sampling (Fig.~\ref{vessel256}), where it preserves substantially more structural details. The CRD offers slightly improved but distorted central resolution and enhancement along preferred orientations. The LLD consistently shows the weakest performance due to strong directional bias and non-uniform sampling density. Under rotational acquisition, the quasi-uniform nature of the FBD becomes more evident, as rotation effectively fills previously unsampled hemispherical regions, enabling sparse FBD to approach the performance of dense static configurations. In contrast, LLD and CRD retain structural gaps after rotation, limiting their improvement. According to prior findings~\cite{hu_location-dependent_2023} that sufficient sampling can avoid spatial aliasing within the FOV, we can expect better imaging quality with the FBD than other distributions.

In several experiments, the differences among evaluation metrics were not pronounced, which may initially appear underwhelming. We attribute this outcome to two primary factors. First, the proposed uniform-spacing sampling criterion effectively approximates near-uniform coverage on the hemispherical surface. Notably, the FBD has been widely regarded as a near-uniform and isotropic sampling scheme, with approximately equal surface area associated with each grid point~\cite{swinbank_fibonacci_2006}. In ultrasound imaging, the Fibonacci (or Fermat’s spiral) arrangement is similarly recognized for its quasi-random angular distribution, which reduces periodicity and mitigates grating lobes~\cite{martinez-graullera_2d_2010,zhang_spiral---curve_2026}. Given that the FBD serves as an established benchmark for quasi-uniform sampling, the comparable performance achieved by the proposed CRD indicates that both designs operate close to the upper bound of sampling uniformity under the given hardware constraints. Second, the relatively large element count (1024) places the system in a high-density sampling regime, where the intrinsic differences among distributions are partially masked by oversampling. As the number of elements decreases, sampling efficiency becomes more critical (Fig.~\ref{vessel256}), resulting in visibly improved structural preservation of the FBD compared with the CRD and the LLD.

Previous ultrasound studies mainly focus on planar arrays and transmit–receive beamforming, particularly with respect to grating lobe suppression and side lobe behavior\cite{ramalli_density-tapered_2015,ramalli_design_2022}. Our work extends the 2D array to a 3D hemispherical array. Although here we only consider purely receiving-mode PA imaging with ideal point sources, owing to acoustic reciprocity, the conclusions regarding sampling uniformity and angular coverage may also extend to ultrasound systems.

Several limitations should be acknowledged. First, although all three distributions were constructed based on a uniform-spacing criterion, this does not guarantee optimality. Sparse array design is a combinatorial optimization problem that is computationally challenging and potentially NP-hard~\cite{ramalli_design_2022}. Second, because of the paper length limit, only a single vascular phantom and a reconstruction method was used for evaluation, which may introduce structure-dependent or algorithm-dependent bias. Third, an identical ESIR was assumed across all elements due to experimental constraints. Although this setup implicitly averages SIR effects over the corresponding incidence angles~\cite{caballero_optoacoustic_2013}, measuring element-specific impulse responses would yield a more realistic system model. Finally, all distributions were constrained to a hemispherical aperture. The effects of reduced cap angles~\cite{choi_deep_2023} or practical illumination ports~\cite{dantuma_fully_2023} require further investigation.

The three distributions generated under the uniform-spacing criterion establish quantitative performance baselines for hemispherical array design. Array configurations that exceed these baselines under comparable constraints may therefore achieve improved sampling efficiency. Based on the comprehensive evaluation across static imaging, noise-perturbed conditions, reduced element counts, and rotational acquisition scenarios, the optimal choice of hemispherical array distribution ultimately depends on the specific imaging objectives and hardware constraints. Overall, the FBD demonstrates superior structural robustness and globally balanced reconstruction quality. In contrast, the CRD provides enhanced resolution and may be advantageous when directional and anisotropic performance is prioritized. Although the LLD exhibits weaker global performance due to non-uniform sampling, it remains the only distribution readily implementable with arc-based arrays, making it attractive under certain fabrication constraints; however, careful design based on the uniform-spacing criterion is required to mitigate its sampling imbalance. More broadly, this work establishes a quantitative framework for hemispherical array distribution design that can be applied during the pre-production stage of system development. By analyzing distribution uniformity and periodicity prior to fabrication, researchers can anticipate resolution characteristics, artifact behavior, and field-of-view coverage, enabling informed design decisions and improving the efficiency of next-generation 3D PACT system development.

\section*{Acknowledgment}
This work was supported by the United States National Institutes of Health grant R00 EB035645. 

\bibliographystyle{ieeetr}
\bibliography{ref}

\end{document}